# Driver Drowsiness Detection with Commercial EEG Headsets

Qazal Rezaee
*Department of Electrical Engineering, K.N. Toosi of Technology*
Tehran, Iran
q.rezaee@email.kntu.ac.ir

Mehdi Delrobaei
*Department of Electrical Engineering, K.N. Toosi of Technology*
Tehran, Iran
delrobaei@kntu.ac.ir

Ashkan Giveki
*Department of Electrical Engineering, University of Tehran*
Tehran, Iran
giveki.ashkan@ut.ac.ir

Nasireh Dayarian
*Department of Electrical Engineering, K.N. Toosi of Technology*
Tehran, Iran
nasireh_dayarian@ee.kntu.ac.ir

Sahar Javaher Haghighi
*Frequency (freqqquency.com)*
Tehran, Iran
s.haghighi@freqqquency.com

***Abstract*— Driver Drowsiness is one of the leading causes of road accidents. Electroencephalography (EEG) is highly affected by drowsiness; hence, EEG-based methods detect drowsiness with the highest accuracy. Developments in manufacturing dry electrodes and headsets have made recording EEG more convenient. Vehicle-based features used for detecting drowsiness are easy to capture but do not have the best performance. In this paper, we investigated the performance of EEG signals recorded in 4 channels with commercial headsets against the vehicle-based technique in drowsiness detection. We recorded EEG signals of 50 volunteers driving a simulator in drowsy and alert states by commercial devices. The observer rating of drowsiness method was used to determine the drowsiness level of the subjects. The meaningful separation of vehicle-based features, recorded by the simulator, and EEG-based features of the two states of drowsiness and alertness have been investigated. The comparison results indicated that the EEG-based features are separated with lower p-values than the vehicle-based ones in the two states. It is concluded that EEG headsets can be feasible alternatives with better performance compared to vehicle-based methods for detecting drowsiness.**

## I. Introduction

Numerous studies have identified driver drowsiness as one of the leading causes of road accidents [1, 2]. Devastating crashes result in high fatality rates since drowsy drivers lose their ability to recognize danger and act accordingly [3]. Fuletra *et al.* have claimed that 30% of all road accidents happen due to driver drowsiness [1]. Driver drowsiness is often caused by four main factors: sleep, work, time, and individual fitness. Most people suffer from a lack of sleep due to work overload resulting in staying awake by consuming caffeine or other stimulants. The lack of sleep increases over several days until the body can no longer withstand this stress, and the person falls asleep involuntarily. The time of occurrence also affects this phenomenon. According to its biological clock, the human brain considers some hours of the day as sleeping time, which is often associated with sunrise and sunset. For example, the brain considers the time between 2 am and 6 am as sleeping time. Prolonged waking time is harmful to the body. Some people use drugs that induce drowsiness or have physical diseases that cause drowsiness. Obesity, physical weakness, or emotional stress can also cause drowsiness. Monitoring the driver before the onset of drowsiness and alerting them in the right time frame is a solution to prevent such drowsiness-related events [4]. Hence, drowsiness detection techniques have gained significant attention in recent years. Driver drowsiness detection techniques are divided into three main categories:
(1) psychological methods, (2) video-based methods (3) and physiological methods.

In psychological methods, questionnaires or tests are used to detect the level of sleepiness. These methods are time-consuming and have a direct relationship with the individual's assessment of their sleepiness. Therefore, these approaches are unsuitable for diagnosing driver drowsiness online, and their results cannot be relied upon.

Video methods are divided into two sub-categories: 1) Behavior-based methods, in which the driver's behavior, such as yawning, closing eyes, blinking, and head position, is monitored through the camera, and the driver is warned if signs of drowsiness are detected. However, since the symptoms of sleepiness vary from person to person, these methods are less accurate. 2) In vehicle-based methods, indicators such as deviation from the line, the movement of the steering wheel, and the amount of pressure on the gas pedal are continuously monitored by sensors embedded in the car. Any change in these cases above a certain threshold indicates increased drowsiness of the driver. Despite their nonintrusive nature and easy recording process, vehicle- based measures suffer from the following major issues that results in low efficiency [4, 7]: (1) Driver drowsiness may not affect vehicle-based parameters in some drivers or until the very severe stages of drowsiness. (2) Vehicle-based parameters depend on external disturbances, such as strong wind and rutted road surfaces, that may interfere with the detection process.

Biosignal-based technologies use physiological signals such as electroencephalography (EEG), electrocardiography (ECG), and electrooculography (EOG) for monitoring and detection purposes [3, 5, 7]. Among all the drowsiness detection measures, EEG is strongly affected by drowsiness and is proven to result in higher accuracy [4, 8]. EEG is suitable for diagnosing drowsiness due to its time resolution and high sensitivity. However, despite the reliable performance of EEG-based approaches, their intrusive nature makes them impractical for driving. The clinical devices used in most cases do not provide the possibility of commercialization of this method due to the disturbance in the individual's driving.

A solution to this issue is using commercial EEG recording headsets for data collection [4, 9, 10]. Advances in dry electrodes and commercial headbands have made it possible for the EEG data acquisition process to take place more quickly. Studies show that these headbands have the potential to record the changes made to the brain signal during sleepiness [4, 9, 12, 13].

In this paper, our goal is to determine whether the performance of EEG-based techniques with consumer-grade headsets is superior to vehicle-based ones. To this end, the EEG signals of sleep-deprived participants were recorded by commercial devices while driving in a simulator. Statistical analyses were performed on the simulator's log data and the brain signals to identify an effective detection technique.

## II. Methodology

### A. *Data Acquisition*

The EEG signals were recorded with commercial headsets (from 4 channels) from 50 participants while driving in a driving simulator. The subjects were fully informed about the test and signed written consent forms before the experiment. All subjects received a small token of appreciation for their participation. Since the simulator's log was unavailable for 2 of the 50 subjects, data analysis was conducted on the remaining 48 series of measurements. Test protocols have been defined based on [4, 11].

Participants should have met the following inclusion criteria: 1) Age: 20-50, 2) Acquisition of a driving license and at least two years of driving experience. Subjects were excluded from the study if they had met the following exclusion criteria: 1) a Body Mass Index (BMI) of more than 40, 2) a sign of sleep disorders or motion sickness, and 3) a history of significant head injury or neurological disorder.

Volunteers were asked to fill out three questionnaires: 1) a general information form asking about their age, height, weight, driving experience, and medical history, 2) Pittsburgh Sleep Quality Index (PSQI): regarding sleep quality, and 3) Epworth Sleepiness Scale (ESS): questions about the possibility of snoozing in different situations.

Participants were encouraged to sleep less (half their regular night's sleep) the night before the experiment (Avg = 4.5 hours). They were also given a heavy meal before the test to make them more prone to drowsiness. The test took place in a dark room, and the subjects were asked to drive in the automatic gearbox mode to prevent any distractions from interfering with the drowsiness process. The duration of the experiment depended on the participant's performance. The experiment would come to an end if one of the following conditions were met: 1) Seventy-five minutes had passed since the beginning of the test, 2) The driver was not able to control the vehicle for any reason, 3) Three alert-to-drowsy cycles were achieved, 4) The driver became restless or requested to stop the test for any other reasons.

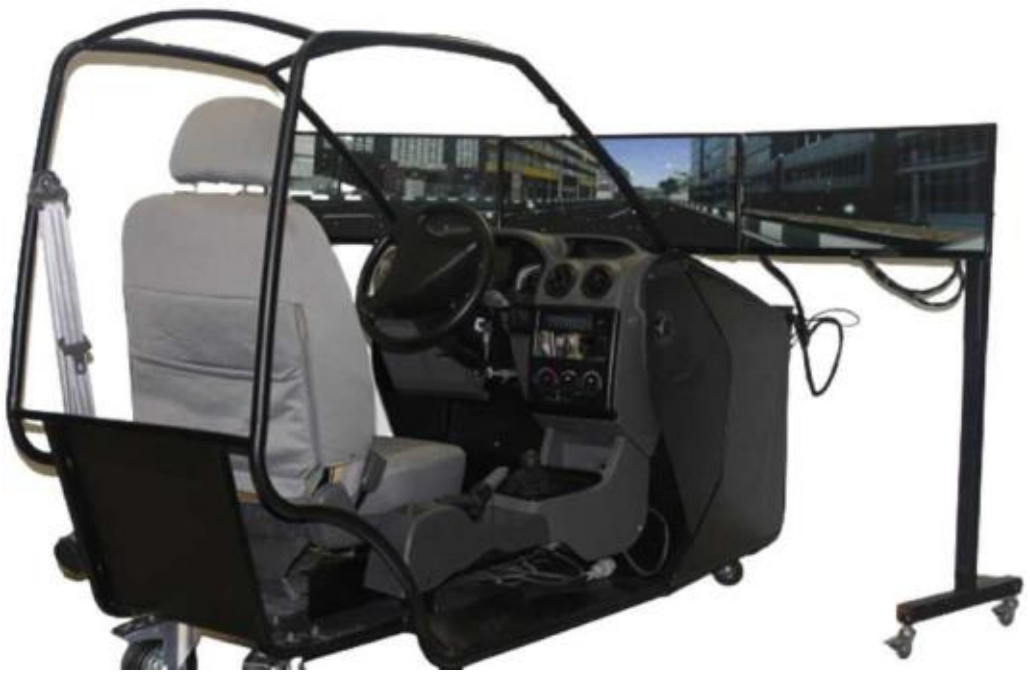

Fig. 1. Nasir Driving Simulator.

The experimental setup included a fixed-base driving simulator (Nasir Driving Simulator, K. N. Toosi University of Technology). Three large LCDs were placed in front of the windshield, almost covering the driver's entire field of vision (Fig. 1). The screens also provided front and side mirrors to complete the subject's view of the road. The simulated road selected for the study had a high fatality rate in accidents due to its almost nonexistent visual attractions. The simulator provided a preprocessed log with the following data derived from its various sensors with a sampling frequency of 30 Hz:

Car position and orientation in x, y, and z axes, car speed, steer, turn state, brake, throttle, clutch, gear, rpm, hand brake, horn, turn signal, flasher, seatbelt, switch state, wiper state, lights state, force sensor, steering angle and speed, lane deviation, lane number, self-aligning torque, torque sensor, steer command, road distance, and trigger event.

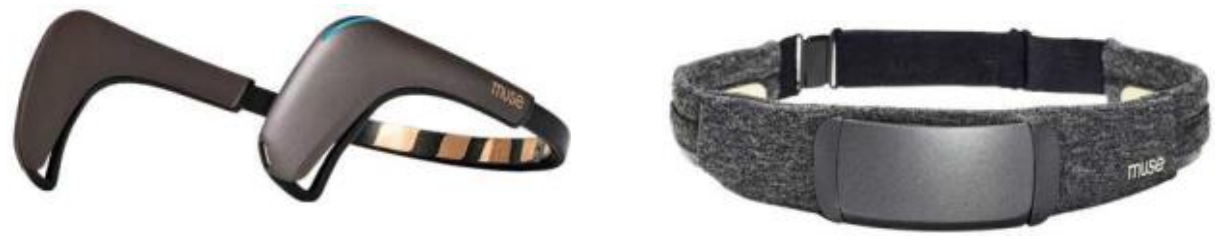

(a) (b)

Fig. 2. The commercial headsets, (a) Muse 2; (b) Muse S.

The commercial headsets used for recording brain signals were Muse 2 and Muse S (Fig 2) [12, 13]. The two devices had similar specifications. Both headsets had four dry electrodes and one reference: AF7, AF8, TP9, and TP10, consistent with the 10-20 system. As in Fig. 3, in the 10-20 system, the electrodes above the eyes are AF7 and AF8, and the electrodes close to the ears are TP9 and TP10. The device is designed to fit around the head, and electrodes are embedded in its fabric headband and positioned near the eyes and ears. The middle electrode, which has a similar position as the FpZ in the 10-20 system, serves as the reference electrode. Besides EEG electrodes, both devices have gyroscope, accelerometer, and PPG sensors that provide the user with four types of data. The sampling frequency rate is 256 Hz.

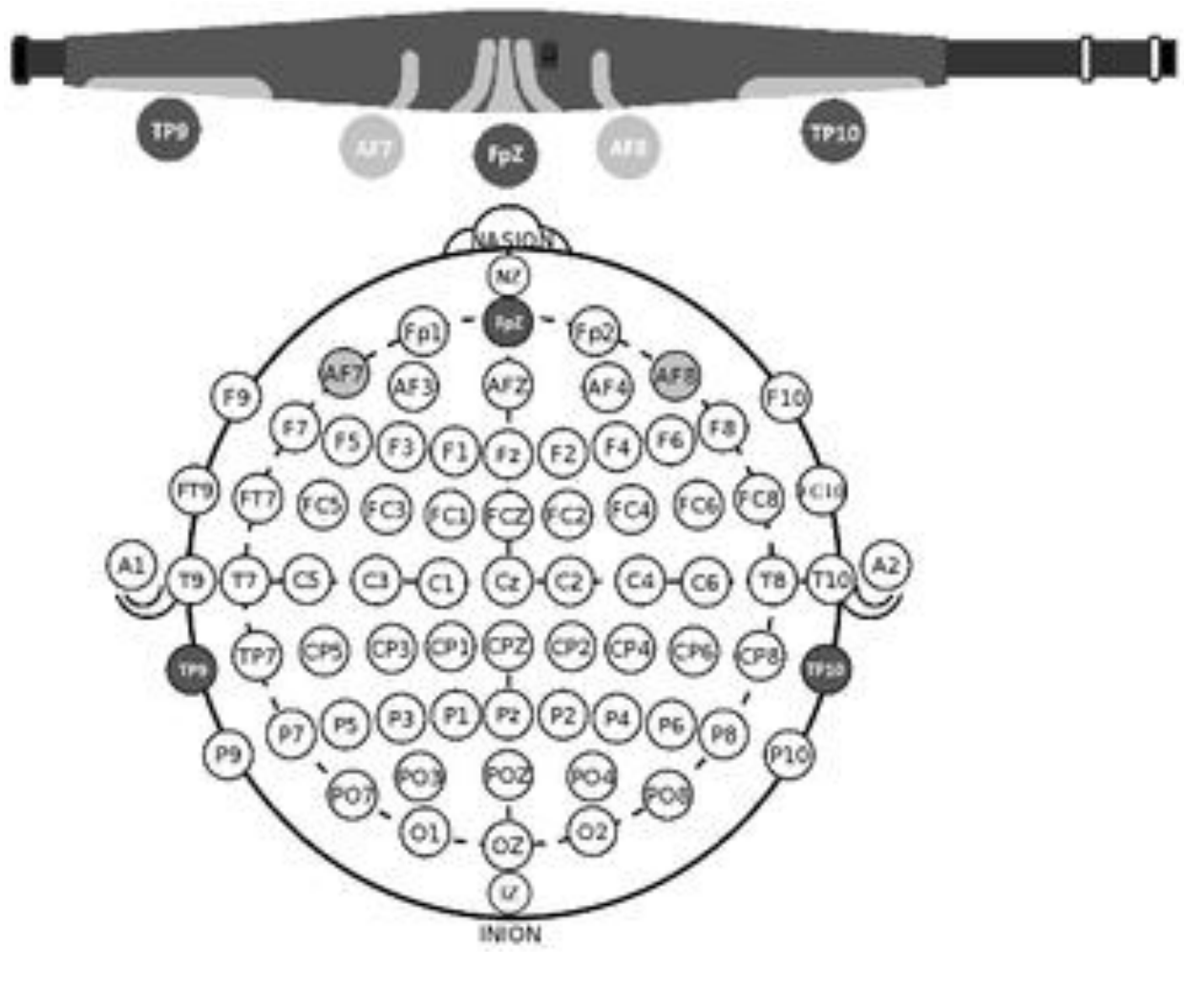

Fig. 3. Muse electrode placement based on the 10-20 system; AF7 and A8F electrodes are located above the eyes, and the TP9 and TP10 are close to the ears. FpZ is the reference electrode placed in the middle of the forehead.

The experiment used a camera to record the participant's face for rating drowsiness and labeling the data. The camera was located on the left side of the steering wheel, capturing the driver's face and neck while not disrupting their view of the screen (Fig 4).

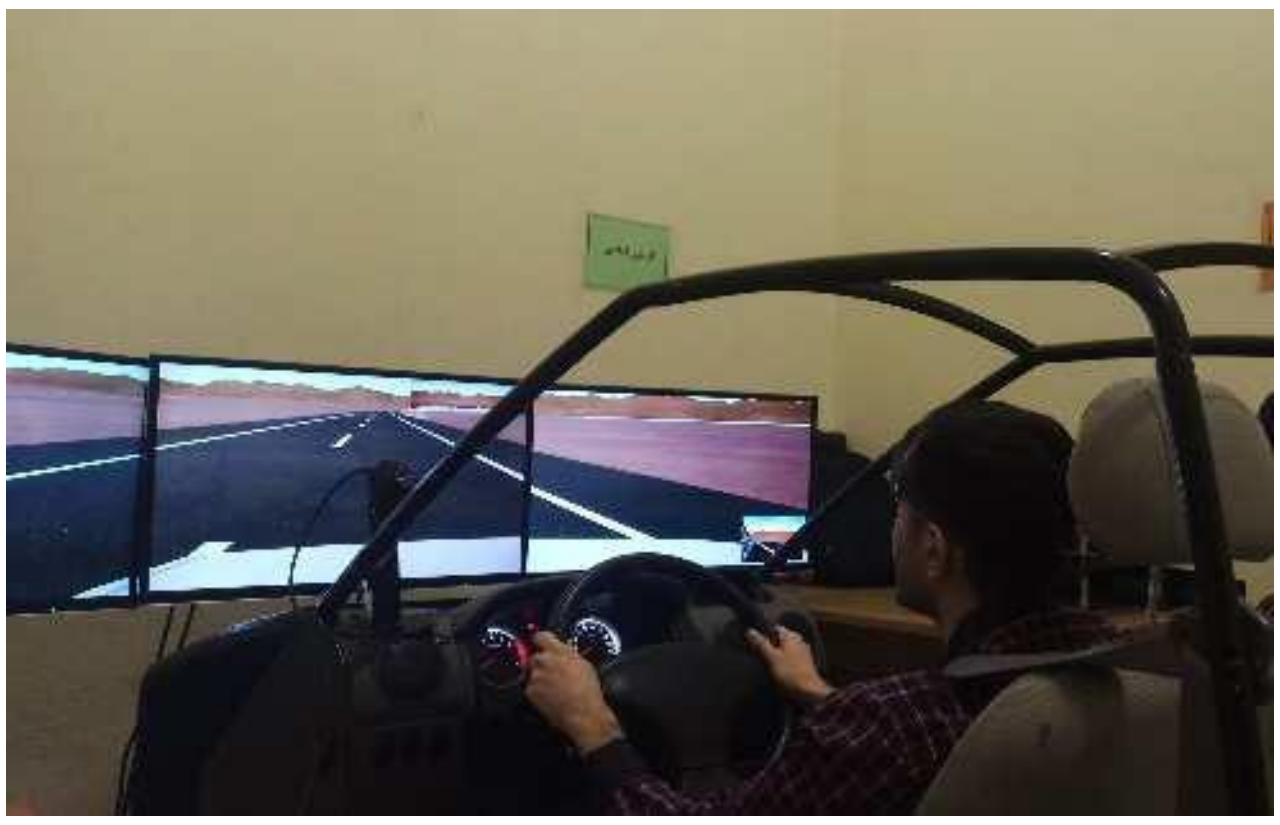

Fig. 4. Simulator during the experiment; The camera located on the left side of the steering wheel captures the driver's face and neck while not disrupting their view of the screen for the driver

## B. *Labeling*

Observer Rating of Drowsiness (ORD) is a subjective method employed in this paper for rating drowsiness and labeling signals. The Drowsiness state of the individual is determined via the video captured from their face and neck during the experiment. Three observers assessed the facial expressions and behavior of the driver in the video (such as eye-lid closure rate, staring, yawning, stretching, and head dropping). The observers score the driver's drowsiness level from 1 (Not drowsy) to 5 (extremely drowsy) every 30 seconds based on the ORD checklist shown in Fig 5 [14, 15]. The final label is then determined by voting among the rating of the three observers. In our study, levels 1 and 2 were considered alert, and levels 3-5 were considered drowsy.

| |
|---|
| **Not Drowsy:** A driver who is not drowsy while driving will exhibit behaviors such that the appearance of alertness will be present. For example, normal facial tone, normal fast eye blinks, and short ordinary glances may be observed. Occasional body movements and gestures may occur. |
| **Slightly Drowsy:** A driver who is slightly drowsy while driving may not look as sharp or alert as a driver who is not drowsy. Glances may be a little longer and eye blinks may not be as fast. Nevertheless, the driver is still sufficiently alert to be able to drive. |
| **Moderately Drowsy:** As a driver becomes moderately drowsy, various behaviors may be exhibited. These behaviors, called mannerisms, may include rubbing the face or eyes, scratching, facial contortions, and moving restlessly in the seat, among others. These actions can be thought of as countermeasures to drowsiness. They occur during the intermediate stages of drowsiness. Not all individuals exhibit mannerisms during intermediate stages. Some individuals appear more subdued, they may have slower closures, their facial tone may decrease, they may have a glassy-eyed appearance, and they may stare at a fixed position. |
| **Very Drowsy:** As a driver becomes very drowsy, eyelid closures of 2 to 3 seconds or longer usually occur. This is often accompanied by a rolling upward or sideways movement of the eyes themselves. The individual may also appear not to be focusing the eyes properly, or may exhibit a cross-eyed (lack of proper vergence) look. Facial tone will probably have decreased. Very drowsy drivers may also exhibit a lack of apparent activity, and there may be large isolated (or punctuating) movements, such as providing a large correction to steering or reorienting the head from a leaning or tilted position. |
| **Extremely Drowsy:** Drivers who are extremely drowsy are falling asleep and usually exhibit prolonged eyelid closures (4 seconds or more) and similar prolonged periods of lack of activity. There may be large punctuated movements as they transition in and out of intervals of dozing. |

Fig. 5. The ORD checklist; Descriptions of progressive drowsiness levels

# III. Result and Discussion

## A. *Drowsiness detection Analysis*

In order to present a comparison between the vehicle-based detection techniques and the commercial EEG-based ones, the two groups of data were analyzed separately. Each group was split into two parts of alert and drowsy data based on the ORD labels. The group with a better separation of the two states performs better detecting drowsiness. The criteria used to determine the performance of each method is the level of marginal significance between the alert and drowsy data. P-value is the probability representing how much the separation of the two states has occurred randomly. A p-value of less than 0.05 is typically considered to be statistically significant. Here, p-values were obtained from the drowsy-related features of each method's alert and drowsy data.

*1) EEG-based analysis:* Since EEG is a noisy signal, several denoising methods were employed to remove noise and artifacts and obtain the signal containing valuable data. The preprocessing steps are as follows:

*a) Epoching:* First, raw EEG signals were epoched into 30-second sections consistent with the ORD labeling intervals.

*Filtering:* In this step, each epoch was filtered with low and high-pass FIR filters with 0.1 and 40 Hz cut-off frequencies, respectively.

*b) Denoising*: Epochs containing more than 30% outliers (data points out of the threshold range of ±70 μv) were considered noisy and omitted from the subject's epochs.

Table 1 represents the number of alert, drowsy, and total epochs of all 48 subjects pre- and post-denoising.

TABLE I. Number of alert, drowsy, and total epochs of all subjects pre- and post-denoising

| | Number of Alert Epochs | Number of Drowsy Epochs | Total Number of Epochs |
|---|---|---|---|
| **Pre-Denoising** | 1058 | 2986 | 4044 |
| **Post-Denoising** | 998 | 2814 | 3812 |

As a result of the denoising process, 5.73% of total epochs, most of which were drowsy, were considered noisy and, therefore, cast aside.

Absolute power spectral density (PSD) and relative PSD (the ratio of the PSD of a single frequency band to the total frequency band) of the frequency bands delta, theta, alpha, beta, and gamma were extracted from the processed EEG signals. Since the signals were recorded by four channels, the total number of features extracted from each epoch was 40. The frequency domain signal was obtained by applying a 1024 point fast Fourier transform (FFT) on each epoch. The absolute and relative PSD of each band was calculated after splitting EEG into the following five frequencies: Delta (0.1- 4 Hz), theta (4-8 Hz), alpha (8-13 Hz), beta (13-30 Hz), and gamma (30-40 Hz).

In order to find features indicating a meaningful separation between the alert and drowsy states, the meaningful separation of each feature in the two states was checked and the p-value was computed. First, each feature's distribution was evaluated by Kolmogorov-Smirnov test. Since some features did not have Gaussian distribution, the p-value was obtained from Wilcoxon rank-sum test. Meaningful features were considered to be the ones with a p- value lower than 0.05. A low p-value means that the feature can separate the two states significantly. The following table shows the p-value of all 40 features of the 48 subjects.

TABLE II. P-VALUES OF EEG FEATURES[A,B]

| *Channel* | *TP9* | *AF7* | *AF8* | *TP10* |
|---|---|---|---|---|
| ***Delta*** | 6.3726e-07 | 0.4024 | 1.5985e-07 | 6.1958e-05 |
| ***Theta*** | 6.6807e-40 | 3.9399e-05 | 4.8593e-17 | 1.9819e-31 |
| ***Alpha*** | 1.3984e-28 | 3.9369e-16 | 3.8816e-19 | 6.2720e-32 |
| ***Beta*** | 2.3431e-38 | 9.2009e-19 | 7.8557e-17 | 1.2218e-23 |
| ***Gamma*** | 5.5560e-35 | 6.3956e-15 | 2.8017e-05 | 1.4233e-24 |

a. Features: absolute PSD of EEG frequency bands of the 4 channels

b. Highlighted features have a p-value of above 0.05

TABLE III. P-VALUES OF EEG FEATURES[C,D]

| *Channel* | *TP9* | *AF7* | *AF8* | *TP10* |
|---|---|---|---|---|
| ***Delta*** | 0.6940 | 0.8034 | 0.0393 | 0.4233 |
| ***Theta*** | 4.3019e-10 | 0.6521 | 0.0183 | 5.6652e-14 |
| ***Alpha*** | 0.8237 | 0.8007 | 5.4517e-05 | 0.0015 |
| ***Beta*** | 0.0167 | 0.0024 | 0.0090 | 0.1401 |
| ***Gamma*** | 0.0097 | 0.0059 | 2.5392e-06 | 0.6121 |

c. Features: relative PSD of EEG frequency bands of the four channels

d. Highlighted features have a p-value of above 0.05

As indicated in tables 2 and 3, the nine highlighted features have p-values greater than 0.05 and should be excluded since the separation of the two states can be considered to have occurred randomly in these features. In absolute PSD features, only 5% have a p-value above 0.05; in relative PSD features, the percentage is 40. The p-values obtained from the former are much smaller, indicating that absolute PSD distinguishes drowsiness and alertness better. Among the five frequency bands, beta, theta, and gamma seem to reflect drowsiness the best, in contrast to delta, which has the lowest number of meaningful features. The results are consistence with the literature claims [16].

*2) Vehicle-based Analysis:* Steer Angle, Steer Speed, Lane Deviation, and Torque Sensor are the vehicle features analyzed in this paper since they are commonly used in previous studies [5, 6]. The average value of the features was calculated for each ORD interval to synchronize the two methods for further comparison. The p-value for each feature was obtained similarly to the EEG-based technique. The following table indicates the p-value of the features in all 48 subjects.

TABLE IV. P-VALUES OF VEHICLE-BASED FEATURES[A,B]

| **Feature** | **Steer Angle** | **Steer Speed** | **Lane Deviation** | **Torque Sensor** |
|---|---|---|---|---|
| **P-value** | 1.8439e-14 | 0.3590 | 0.0075 | 2.9897e-09 |

a. Features: steer angle and speed, lane deviation, and torque sensor

b. Highlighted features have a p-value of above 0.05

As presented, the vehicle-based method's four features, steer angle, lane deviation, and torque sensor can differentiate drowsiness from alertness. The steer speed's p-value is more significant than 0.05 and, thus, should be excluded.

## IV. CONCLUSIONS

This work compared the vehicle-based and the commercial EEG-based drowsiness detection methods. Each method has its own merits and demerits. Vehicle-based measures have a nonintrusive nature, so the data collection process with these techniques is easy. However, it is challenging to obtain high accuracy in detecting drowsiness using these approaches since they are highly affected by external disturbances, and drowsiness might not affect vehicle-based parameters in less severe levels of drowsiness. On the other hand, EEG-based measures have proven high accuracy because drowsiness directly affects brain signals. The main issue with these methods is the complexity of their data acquisition process. Many studies have utilized clinical devices in this regard, and although the results seem compelling, they do not outweigh the difficulty of the collection process. In our case study, we used commercial devices to substitute the clinical ones, making the signal recording much easier and less costly. Our goal was to investigate whether this recording process performs better than vehicle-based approaches. To assess the performance of each method, a comparison was made between the p-values of each method's features., It can be concluded that EEG features acquired by commercial headsets can distinguish drowsiness and alertness better than vehicle-based features.

## ACKNOWLEDGMENT

This research was financially supported by the Frequency Group, Bonda Co. Test protocols were developed with the insight and expertise of our colleagues in the company.

We thank the Virtual Reality Laboratory of the Faculty of Mechanic Engineering of K.N. Toosi University, Tehran, Iran, for providing the driving simulator.